# "PROCESS": a systems code for fusion power plants - Part 2: Engineering

M Kovari*, F. Fox, C. Harrington, R. Kembleton, P. Knight, H. Lux, J. Morris

CCFE, Culham Science Centre, Abingdon, Oxon, OX14 3DB, UK
*Corresponding author.  Tel.: +44 (0)1235-46-6427.  E-mail address: michael.kovari@ccfe.ac.uk

**Abstract**
PROCESS is a reactor systems code – it assesses the engineering and economic viability of a hypothetical fusion power station using simple models of all parts of a reactor system.  PROCESS allows the user to choose which constraints to impose and which to ignore, so when evaluating the results it is vital to study the list of constraints used.  New algorithms submitted by collaborators can be incorporated – for example safety, first wall erosion, and fatigue life will be crucial and are not yet taken into account.  This paper describes algorithms relating to the engineering aspects of the plant.  The toroidal field (TF) coils and the central solenoid are assumed by default to be wound from niobium-tin superconductor with the same properties as the ITER conductors.  The winding temperature and induced voltage during a quench provide a limit on the current density in the TF coils.  Upper limits are placed on the stresses in the structural materials of the TF coil, using a simple two-layer model of the inboard leg of the coil.  The thermal efficiency of the plant can be estimated using the maximum coolant temperature, and the capacity factor is derived from estimates of the planned and unplanned downtime, and the duty cycle if the reactor is pulsed.  An example of a pulsed power plant is given.  The need for a large central solenoid to induce most of the plasma current, and physics assumptions that are conservative compared to some other studies, result in a large machine, with a cryostat 36 m in diameter.  Multiple constraints, working together, restrict the parameter space of the optimised model.  For example, even when the ratio of operating current to critical current in the TF coils is increased by a factor of five, the total coil cross-section decreases only a little, because of the need for copper stabiliser, insulation, and structural support.  The result is that the plasma major radius hardly changes.  It is these surprising results that justify the development of systems codes.

1. **Introduction`**

While physicists at experimental machines investigate whether a fusion plasma can be confined, it is equally important to assess whether a fusion plant is feasible from the engineering and economic points of view.  Information on this is collated in reactor systems codes, which contain simple models of an entire power plant, including physics, engineering and costs.  The PROCESS systems code has been used for many years, and details of its physics algorithms and general structure have been published previously (1).  This paper describes the engineering assumptions and models.

PROCESS is one of the most flexible of all reactor systems codes.  It finds a set of parameters that maximise (or minimise) a Figure of Merit chosen by the user, while being consistent with the constraints, by adjusting a set of variables known as iteration variables.  Both the constraints and the iteration variables are chosen by the user from an extensive selection.  In effect, therefore, the user can choose which are input variables and which are outputs.  Only those constraints specified by the user are enforced.  We describe PROCESS version 393.

Sections 2 to 8 describe the models for the superconducting magnets, the first wall, blanket and shield, the flow of thermal power and its conversion to electricity, and the availability model. Section 9 describes a pulsed DEMO model obtained using PROCESS, illustrated in Figure 1.

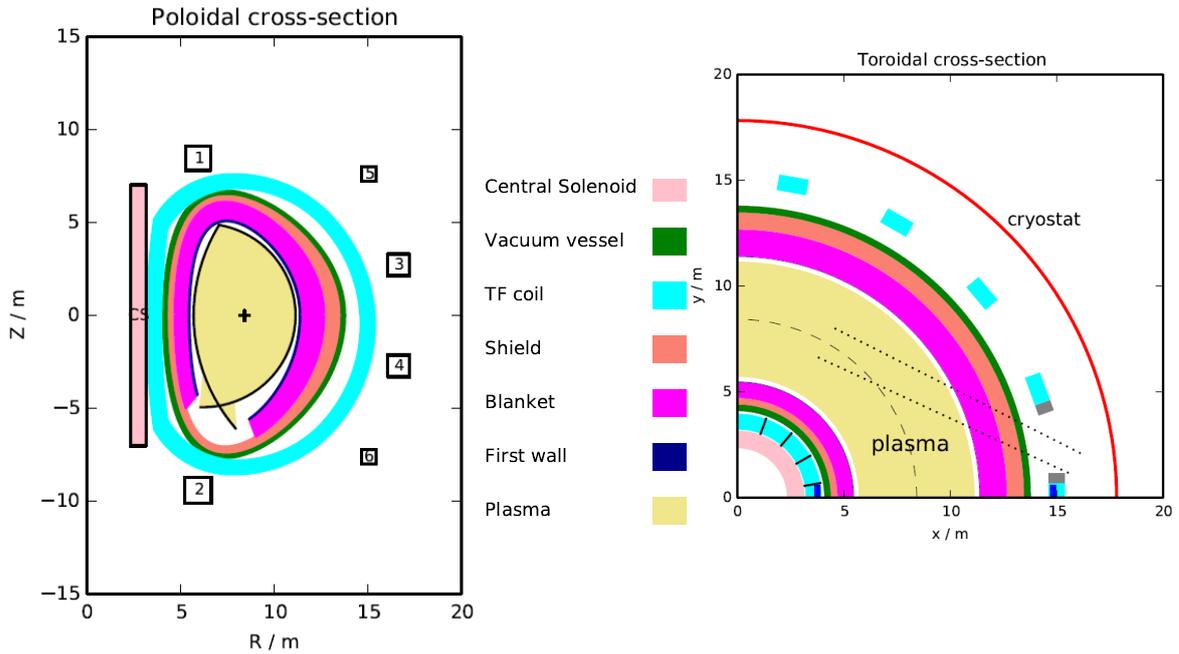

Figure 1. Cross-sections of PROCESS model of a pulsed reactor. In one of the TF coils the winding pack is shown in blue, and the shielding for the neutral beam duct in grey. The thermal shielding which is needed to separate the cold superconducting coils from the hot reactor inside, and from the cryostat outside, is not included explicitly. The ports for diagnostics and remote handling are not shown because they are not modelled in PROCESS.

## 2. Toroidal field coil (TFC)

In PROCESS the TF coil consists of a winding pack with a homogenous current density, surrounded by a structural case. Several conductor models are available, but the default assumes the use of forced-flow helium cooled superconducting cables, such as the cable-in-conduit type. AC losses are not taken into account. A number of constraints are available for the TF coil but, as always, are only enforced if selected by the user. They include (a) stress in case, (b) stress in conduit, (c) ratio of operating current to critical current, (d) superconductor temperature margin, (e) quench voltage, and (f) quench temperature. Details are below.

Numerous input parameters are available as iteration variables, in which case PROCESS will vary them automatically, making them effectively outputs. These include the toroidal field on axis, the radial thickness of the TFC, the coil current per turn, the copper fraction in the conductor, the overall current density in TF coil inboard legs, and many others.

The TFC is symmetrical, each half being approximated by 4 circular arcs along the edge facing the plasma. The height is determined purely by the vertical build – the coil is not required to have a constant tension "D" shape. Note that the inboard leg is not exactly straight. This model is used only to calculate the mass, inductance and stored energy.

### 2.1. Access required for neutral beams

The maximum tangency radius for the neutral beams is determined by the size and shape of the TF coils, as the beams need to pass between them at an angle. This may be an important constraint on the achievable neutral beam current drive, as this is usually maximum when the tangency radius is equal to or slightly greater than the major radius (1),(2). Figure 2 shows the geometry and symbols used. The need for remote handling may impose additional constraints. If the blanket modules run the full height of the machine, and are accessed for maintenance from above, then it would not be acceptable for a neutral beam duct to cut the whole blanket module in half, but this constraint has not been included.

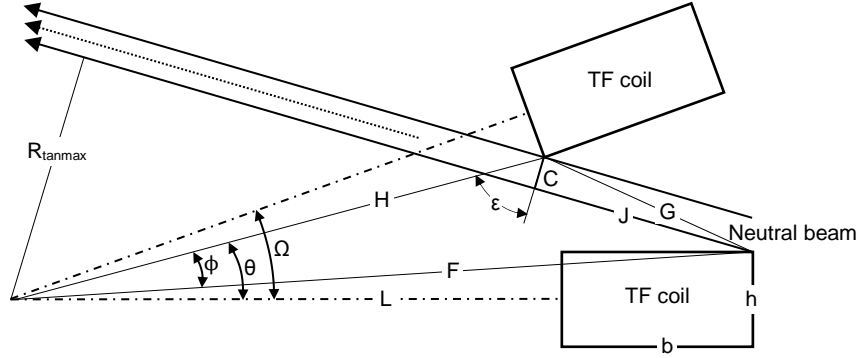

**Figure 2.** Geometry for neutral beam access between TF coils

$$\Omega = \frac{2\pi}{N_{TF}} \qquad 1$$

$$F = \sqrt{\left(\frac{h}{2}\right)^2 + (L+b)^2} \qquad 2$$

$$H = \sqrt{L^2 + \left(\frac{h}{2}\right)^2} \qquad 3$$

$$\theta = \Omega - \tan^{-1}\left(\frac{h/2}{L}\right) \qquad 4$$

$$\phi = \theta - \tan^{-1}\left(\frac{h/2}{L+b}\right) \qquad 5$$

$$G = \sqrt{H^2 + F^2 - 2HF\cos\phi} \qquad 6$$

$N_{TF}$ is the number of TF coils, and $C$ is the width required for the neutral beam duct, including any shielding required to protect the TF coils on either side, the thickness of the duct wall, the thermal shields and the vacuum gaps.

$$J = \sqrt{G^2 - C^2} \qquad 7$$

$$\epsilon = \sin^{-1}\left(\frac{F\sin\phi}{G}\right) - \tan^{-1}\left(\frac{J}{C}\right) \qquad 8$$

The maximum possible tangency radius, which can be applied as a constraint, is

$$R_{tanmax} = H\cos(\epsilon) - \frac{C}{2} \qquad 9$$

## 2.2. TF coil current density

In the default model the parameterisation of critical current density in Nb$_3$Sn as a function of magnetic field $B$, temperature $T$ and strain $\varepsilon$ uses the ITER formulation (3), correcting for the strand cross-section and the fraction of the strand occupied by copper (Fig. 3). The fitting parameters are in Table 1. The peak magnetic field is calculated using a parametric fit to detailed calculations from the Biot-Savart law, while the temperature and strain are set by the user.

Critical strand current in the reference strand (0.82 mm diameter, copper : non-copper ratio=1) :

$$I_C = \frac{C}{B} s(\epsilon)(1-t^{1.52})(1-t^2)b^p(1-b)^q \qquad 10$$

Critical temperature:

$$T_C^*(B,\epsilon) = T_{C0max}[s(\epsilon)]^{1/3}(1-b_0)^{1/1.52} \qquad 11$$

Critical field:

$$B_{C2}^*(T,\epsilon) = B_{C20max} s(\epsilon)(1-t^{1.52}) \qquad 12$$

Strain function:

$$s(\epsilon) = 1 + \frac{1}{1-C_{a1}\epsilon_{0,a}}\left[C_{a1}\left(\sqrt{\epsilon_{sh}^2 + \epsilon_{0,a}^2} - \sqrt{(\epsilon-\epsilon_{sh})^2 + \epsilon_{0,a}^2}\right) - C_{a2}\epsilon\right] \qquad 13$$

$$\epsilon_{sh} = \frac{C_{a2}\epsilon_{0,a}}{\sqrt{C_{a1}^2 - C_{a2}^2}} \qquad 14$$

Reduced magnetic field:

$$b = \frac{B}{B_{C2}^*(T,\epsilon)} \qquad 15$$

Reduced magnetic field at zero temperature:

$$b_0 = \frac{B}{B_{C2}^*(0,\epsilon)} \qquad 16$$

Reduced temperature at zero field:

$$t = \frac{T}{T_C^*(0,\epsilon)} \qquad 17$$

**Table 1**. ITER Reference Scaling Parameters for TF and CS Conductor Design (4)

| | | TF | CS |
|---|---|---|---|
| C | Scaling constant for strand current (AT) | 16500 | 18700 |
| $B_{C20max}$ | upper critical field at zero temperature and strain | 32.97 | 32.57 |
| $T_{C0max}$ | critical temperature at zero field and strain | 16.06 | 17.17 |
| p | low field exponent of the pinning force (p < 1, p ≈ 0.5) | 0.63 | 0.62 |
| q | high field exponent of the pinning force (q ≈ 2) | 2.1 | 2.125 |
| $C_{a1}$ | Strain fitting constant, | 44 | 53 |
| $C_{a2}$ | Strain fitting constant, | 4 | 8 |
| $\epsilon_{0,a}$ | residual strain component | 0.00256 | 0.0097 |

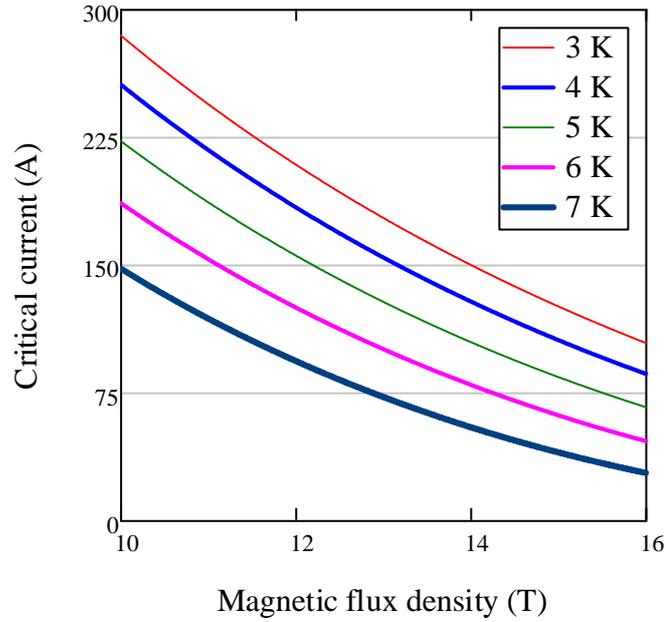

**Figure 3**. Critical current of Nb$_3$Sn TFC strand using ITER parameterization. Strain = -0.3%

The critical current density in a general strand is

$$J_c^{str} = J_c^{sup}(1 - f_{Cu}) \qquad 18$$

where $f_{Cu}$ is the fraction of each strand occupied by copper, and the current density in the superconductor, $J_c^{sup}$, is derived from the reference strand described above. The critical current of the cable is

$$I_c^{cable} = J_c^{str} A_{cs} (1 - f_{He}) \qquad 19$$

where $A_{cs}$ is the interior cross-sectional area of the cable, $f_{He}$ is the fraction of that area occupied by helium coolant. The actual current per turn is an input, and is available as an iteration variable. The temperature margin is the difference between the temperature at which the critical current equals the actual current, and the actual temperature.

The TF conductor is taken to be of the cable-in-conduit design, as illustrated in Figure 4. The cross-sectional area of the conductor is calculated, after allowing for the rounded corners and the fraction occupied by helium coolant.

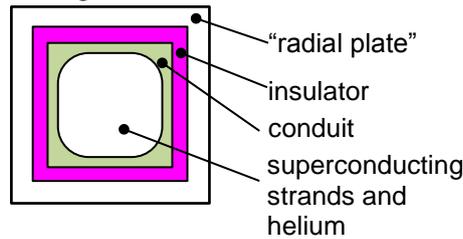

Figure 4. Layout of the TFC cable, used for calculating current density and effective Young's modulus. Additional structural material is taken into account as shown, and is described as "radial plates".

## 2.3. Quench protection of TFC

During a quench the coil needs to be discharged into an external resistor to protect the cable and limit the induced voltage. The maximum permissible winding temperature during a quench provides a limit on the current density. It is assumed that the superconductor, copper and helium remain in thermal equilibrium with each other, but no heat is taken up by the conduit. The variation of heat capacity and resistivity with temperature are taken into account, but not the effect of the magnetic field on the resistivity of the copper stabiliser. The dump resistor has a resistance much higher than that of the coil during the quench. The maximum current density in the cable space is given (5) by

$$J = \sqrt{\frac{VI_{op}}{E_{stoTF}} f_{Cu}(f_{He}I_{He} + f_{Cu}I_{Cu} + f_{sc}I_{sc})} \qquad 20$$

where V is the peak voltage developed across the coil, $I_{op}$ = current per turn, $E_{stoTF}$ = stored energy per coil, $f_{Cu}$, $f_{He}$, $f_{sc}$ are the volume fractions of helium, copper and superconductor in the cable space, and

$$I_{He} = \int_{T_0}^{T_{max}} \frac{\rho_{He} C_{He}}{\eta} dT \qquad 21$$

$$I_{Cu} = \int_{T_0}^{T_{max}} \frac{\rho_{Cu} C_{Cu}}{\eta} dT \qquad 22$$

$$I_{sc} = \int_{T_0}^{T_{max}} \frac{\rho_{sc} C_{sc}}{\eta} dT \qquad 23$$

where $\rho$ = density, $C$ = specific heat capacity, $T_0$ is the operating temperature, $T_{max}$ is the maximum temperature reached (input by the user), and $\eta$ = electrical resistivity of copper. No limits are placed on the pressure in the conduit or on the stability of the conductor. The user can specify the time constant for the quench ($t_{dump}$), and can also set an upper limit for the peak voltage V developed by the quench. The stored energy per coil is $LI_{op}^2/(2N_{TF})$, where L is the total inductance of the TF coil set, therefore

$$V = 2 \frac{E_{stoTF}}{t_{dump} I_{op}}. \qquad 24$$

This assumes that the energy deposited in a single dump resistor is derived from the energy stored in a single TF coil only.

### 2.4. Stress in the TFC

The net forces on the TF coils are inward (toward the major axis of the machine). In principle these can be supported on the central solenoid (as on JET) or on an additional structural part known as a bucking cylinder, or the straight sections of the coils can form an arch, also known as a vault (as on ITER), illustrated in Figure 5.

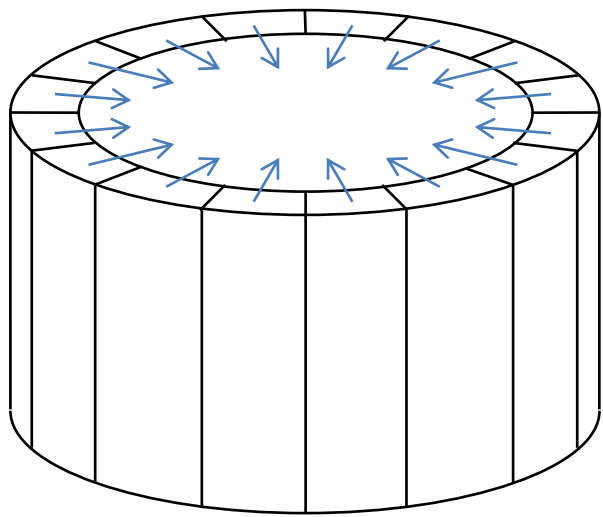

Figure 5. A portion of the inboard part of the TF coils, showing a self-supporting arch or vault. The arrows represent the net Lorentz force on each coil.

Tolerance problems make it difficult to use more than one method. Using the CS to support the TF coils is efficient, but causes problems as the stresses vary during the pulse, so PROCESS assumes that a vault is used.

Only the stresses in the inboard leg at midplane are calculated. The field from the CS and PF coils is not included in any way, and only steady-state stress is calculated. Figure 6 shows the layouts assumed.

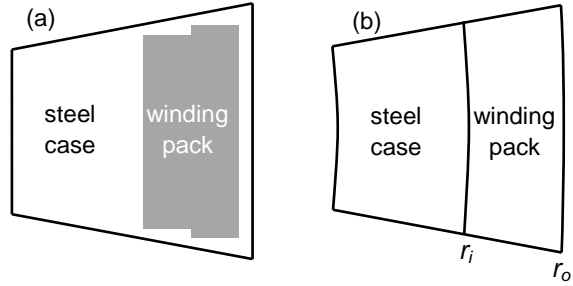

Figure 6. Mid-plane cross-section of inboard leg of TFC. (a) As used for calculating cross-sectional areas, (b) as used for calculating stress, showing the major radii of the dividing surfaces.

The ITER TF coils have steel radial plates that provide extra support for the cable. To represent this in a simplified way, PROCESS allows additional structural material within the winding pack, of the same material as the TF case (see Figure 4). The winding pack is modelled in the stress calculation as a homogeneous material. Because this cable is assumed to be square, an isotropic elastic modulus is used in the horizontal plane. The vertical Young's modulus is used for the separate calculation of vertical tensile stress. The shear stresses with a vertical component are zero because the out-of-plane forces on the coil due to the poloidal field are neglected.

To calculate the effective Young's modulus of the winding pack each turn of the coil is split conceptually into series and parallel parts as shown in Figure 7. While the turn is square and therefore symmetric, there are two different ways to split it, giving different values for the smeared Young's modulus, although the difference is small and has no physical significance. One value has been chosen arbitrarily, as follows. Assuming that the force is applied in the direction of the arrow, the components $a$ to $d$ carry the load in parallel, so they have the same strain. In component $b$, for example, the insulation ($i$) and the radial plate ($rp$) are series, so they have the same stress. Components $b$ and $c$ have very little stiffness, since they are mostly composed of insulation, superconductor and helium.

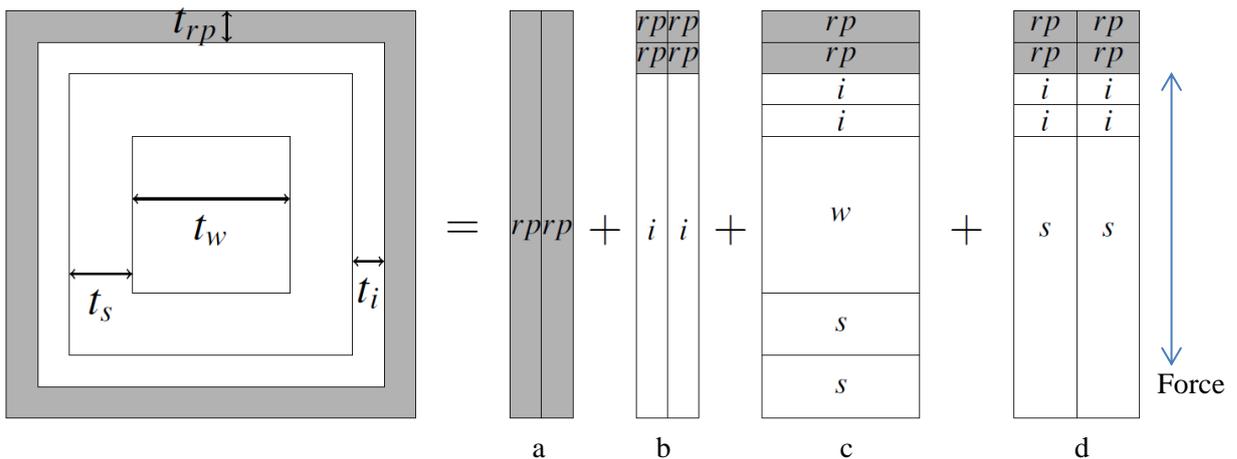

Figure 7. The cable plus radial plate shown split into series and parallel components. The insulation is $i$, conduit $s$, the conductor $w$, and the radial plate $r$.

The effective Young's moduli for the components labelled a – d in Figure 7 are therefore

$$E_a = E_{rp} = E_s \qquad 25$$

$$E_b \approx 0, \quad E_c \approx 0, \qquad (26)$$

$$E_d = \frac{t_{tot}}{\frac{2t_{rp}}{E_{rp}} + \frac{2t_i}{E_i} + \frac{t_w + 2t_s}{E_s}}, \qquad (27)$$

where the subscripts are

| | |
|---|---|
| s | conduit (steel) |
| rp | radial plate (also steel) |
| i | insulator |
| w | conductor |

The combined modulus is

$$E_\beta = \frac{2t_{rp}E_s + 2t_s E_d}{t_{tot}}, \qquad (28)$$

$$t_{tot} = 2(t_{rp} + t_i + t_s) + t_w \qquad (29)$$

When the load is vertical the conductor, conduit, insulator and radial plate are all in parallel, so the modulus is

$$E_z = \frac{1}{t_{tot}^2}\big[E_w t_w^2 + E_s((t_w + 2t_s)^2 - t_w^2) + E_i((t_w + 2t_s + 2t_i)^2 - (t_w + 2t_s)^2) + E_{rp}\big((t_w + 2t_s + 2t_i + 2t_{rp})^2 - (t_w + 2t_s + 2t_i)^2\big)\big] \qquad (30)$$

Poisson's ratio $v$ is taken as 0.3 in all cases.

### 2.4.1. Horizontal stress components

The TF coil is split into two layers: the winding pack and the case (Figure 6). The force on the inboard leg is primarily directed towards the machine axis, so the bulk of the case is best used on the side nearest this axis. The thin layer of steel on the plasma-facing side is ignored. The toroidal field at the outer edge of the winding pack is $B_{maxTF}$.

The Lorentz force per unit volume is radial,

$$F_r = jB. \qquad (31)$$

The current density, assumed to be constant, is

$$j = \frac{I_{TFC}}{\pi(r_0^2 - r_i^2)} \qquad (32)$$

where $r_0$ and $r_i$ are the radii of the winding pack and $I_{TFC}$ is the total current in the set of TF coils. Inside the winding pack at major radius $r$, Ampere's law gives the field, assumed to be toroidally uniform, as

$$B = \frac{\mu_0}{2\pi r} j\pi(r^2 - r_i^2). \qquad (33)$$

By toroidal symmetry, the local displacement $u$ is purely radial. Using a cylindrical co-ordinate system $(r,\theta)$, and the definition of Poisson's ratio $v$, it can be seen that the strain components are

$$\epsilon_r = \frac{du}{dr} = \frac{1}{E}(\sigma_r - v\sigma_\theta) \qquad (34)$$

$$\epsilon_\theta = \frac{u}{r} = \frac{1}{E}(\sigma_\theta - v\sigma_r) \qquad (35)$$

where $v$ is Poisson's ratio, and $E$ is the combined Young's modulus for the layer in question.

From local force balance a differential equation is derived for each layer,

$$\frac{d^2u}{dr^2} + \frac{1}{r}\frac{du}{dr} - \frac{u}{r^2} = -\frac{(1-v^2)}{E_c}F_r = \alpha r + \frac{\beta}{r} \qquad 36$$

where

$$\alpha = \frac{(1-v^2)}{E_c}\frac{\mu_0 j^2}{2} \qquad 37$$

$$\beta = -\alpha r_i^2 \qquad 38$$

The solution for the displacement is:

$$u = C_1 r + \frac{C_2}{r} + \frac{\alpha}{8}r^3 + \frac{\beta}{2}r \log r \qquad 39$$

where $C_1$ and $C_2$ are constants of integration, defined for each layer. The boundary conditions give a set of four equations, which are solved using Gaussian elimination each time the code is run, to give the strain as a function of radius in each layer. The peak tangential stress is much greater than the peak radial stress. The highest stress in the case is at the innermost radius and the highest stress in the winding at the innermost radius of the winding. In the winding pack the averaging procedure is reversed to give the stress in the structural portion (conduit and radial plates) and the strain in the conductor.

### 2.4.2. Vertical stress components

By Ampere's law the vacuum toroidal field at major radius $r$ inside the TF coil is

$$B_v(r) = \frac{\mu_0 I_{TFC}}{2\pi r}. \qquad 40$$

The field outside the coil is nearly zero (exactly zero for an infinite number of coils), so the average field inside the winding pack is approximately half this value. The Lorentz force per unit length of coil is therefore about

$$f = \frac{B_v(r)}{2}\frac{I_{TFC}}{N_{TF}} = \frac{\mu_0 I_{TFC}^2}{4\pi r N_{TF}} \qquad 41$$

($N_{TF}$ = number of TF coils)

This force is perpendicular to the coil segment, and its vertical component is $fdr$ where $dr$ is the component of the segment length along the major radius. The total vertical force on upper half of the coil is then

$$F_z = \int_{R_{TFCin}}^{R_{tot}} f\,dr = \frac{\mu_0 I_{TFC}^2}{4\pi N_{TF}} \ln\left(\frac{R_{tot}}{R_{TFCin}}\right). \qquad 42$$

It can be shown that the tensile force in the inboard leg is half of this. The superconductor is likely to be twisted, which minimises its tensile stress, so we neglect the stiffness of the conductor in the vertical direction. The vertical tensile stress in the inboard leg $\sigma_z$ is given by dividing this force by the total area of case, conduit and radial plate. In the vertical direction the components of the winding pack are in parallel, so the fractional extension of the superconductor is $\sigma_z/E_z$. This is not a true strain if the conductor is twisted.

### 2.4.3. Stress criteria

PROCESS assumes that the principal axes of the stress at the mid-plane are vertical, radial and tangential, so there are no shear stresses in this coordinate system. The von Mises stresses in the case and in the structural part of the winding pack are then given in terms of the radial, tangential and vertical stress components $\sigma_r$, $\sigma_t$, $\sigma_z$. Because the structural material inside the winding pack is assumed to take the form of radial and tangential webs (Figure 7), the limiting von Mises stress in this zone is given by the larger of the two values:

$$\sigma_{vonMises1} = \sqrt{\frac{1}{2}(\sigma_r^2 + (\sigma_r - \sigma_z)^2 + \sigma_z^2)}, \qquad 43$$

$$\sigma_{vonMises2} = \sqrt{\frac{1}{2}(\sigma_t^2 + (\sigma_t - \sigma_z)^2 + \sigma_z^2)}. \qquad 44$$

In both zones the peak von Mises stress always occurs at the inner radius. These peak values can each be constrained to be no more than the permissible value, specified by the user.

### 3. Central solenoid (CS)

The central solenoid provides a loop voltage for plasma initiation and current ramp-up. For a pulsed reactor it also provides some of the voltage required to maintain fusion burn. Quench protection is not taken into account. The CS contains a fraction of steel structural material, whose allowable stresses are the same as for the TF coils (section 2.4.3). Only steady-state stress is considered, with no allowance for fatigue due to cyclic stress, although fatigue is likely to be significant for a pulsed reactor. The hoop force is calculated using the approximation,

$$F_{J \times B} = \frac{B_o + B_i}{2} I_{CS} R_{CS} \qquad 45$$

where $B_o$ and $B_i$ are the fields at the outer and inner edge of the coil (calculated by the code, taking account of the field due to the plasma and all the other PF coils), $I_{CS}$ is the total central solenoid current, and $R_{CS}$ is the radius of the midline of the coil. The minimum cross-section of steel is calculated from the hoop force and the allowable stress. The current density in the superconductor is derived taking account of the area of steel and of helium coolant. The critical current density in the superconductor is calculated using same parameterisation for $Nb_3Sn$ described in 2.2 above, at Beginning of Flat-top and End of Pulse. By comparing the actual current density and the critical current density, the temperature margin is derived at each time point. The smaller of these values is reported, and can be given a lower limit using a constraint.

### 4. Poloidal field (PF) coils

The current per turn (i.e. the current in the conductor) is an input parameter. The number of turns in each coil is then calculated from the total current. The mass of superconductor in each coil is calculated from the cross-section, length, void fraction and density. The tangential tension (hoop) force is

$$F = R \frac{(B_{PF} + B_{PF2})}{2} I \qquad 46$$

where R is the radius of the coil, $B_{PF}$ is the field at the inner edge, $B_{PF2}$ is the field at the outer edge, and I is the peak current. The cross-sectional area and mass of the structural material, assumed to be steel, in each PF coil is calculated using the maximum permissible tensile stress in the steel, and a specified fraction of the hoop force to be supported by the steel. The steel required is not included in the dimensions output.

### 5. First wall, blanket and shield

The neutron wall loading is calculated by dividing the neutron power by the wall area.

$$W_{all} = \frac{P_{neut} V}{F_{area} S_{area}} \qquad 47$$

where $S_{area}$ = plasma surface area, $P_{neut}$ = neutron fusion power per volume, $V$ is the plasma volume and $F_{area}$ = user-specified ratio between first wall area and plasma surface area. An upper limit can be imposed on the neutron wall load, which is a nominal quantity, only loosely representative of the actual neutron flux whose angular and energy spectra are important.

The volumes of the first wall, blanket, shield and vacuum vessel are calculated using one of two models: they may be D-shaped in cross-section (Figure 8), in which the inboard part is cylindrical, or both inboard and outboard parts may be defined by half-ellipses. In practice the thicknesses of the blanket and shield at the top are taken as the mean of the inboard and outboard thicknesses, while the vacuum vessel thickness is constant.

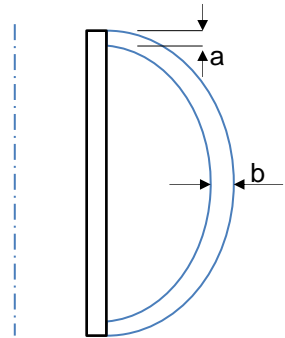

Figure 8. D-shaped model used for calculating volumes of blanket, shield and vacuum vessel: each is half a toroidal shell. The thickness varies between a and b. The inboard part is modelled as a cylinder.

The mass of the blanket is determined by its volume and by its volumetric composition which by default includes titanium beryllide ($TiBe_{12}$), lithium orthosilicate ($Li_4SiO_4$), helium and steel.

## 6. Plant power balance

### 6.1. Reactor power

The plant power flowchart is shown in Figure 9.

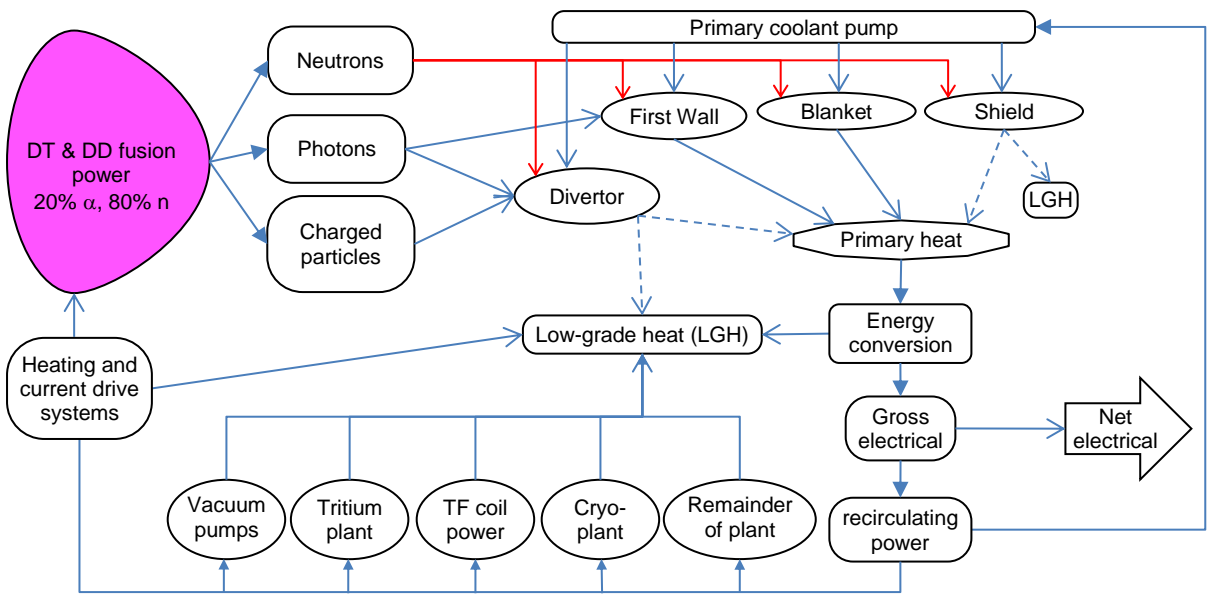

**Figure 9**. Power flows. LGH is low grade heat, rejected to the environment. Red lines represent neutrons. The dashed lines represent alternative options. The heat loads on the cryogenic components are not included in this diagram.

The cooling system consists of two parts – the part heated by "primary" heat, which contributes to electricity production, and the part heated by "low grade heat", which does not. The options for how power is divided are shown in Table 2. The wall-plug efficiency of the heating and current drive system used is specified by the user.

Table 2. Primary heat (useful for electricity production)

| Component | Type of heating | Fraction |
|---|---|---|
| First wall & blanket | nuclear heating photon radiation pumping power | 100% |
| Shield | nuclear heating, pumping power | 0% or 100% |
| Divertor | fusion power to α particles nuclear heating photon radiation pumping power | 0% or 100% |

## 6.2. Power deposition

The deposition of nuclear power in the reactor components are derived from the transport model for neutrons and secondary particles illustrated in Figure 10. (The photon power on the first wall and divertor are derived from the radiation model described in (1).)

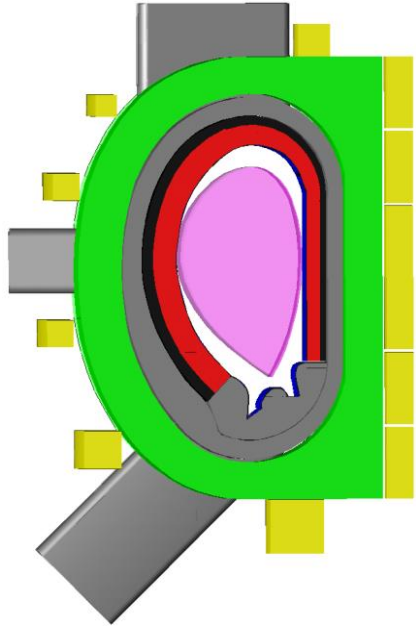

Figure 10. The neutronic model used to derive heat deposition. The plasma is shown in pink, breeding blanket in red, shield in black, vacuum vessel (including ports) in grey, the TF coil in green and the CS and PF coils in yellow. The face of the first wall armour facing the plasma is blue.

The following functions provide a reasonable fit to the results for nuclear heating power in each component, as follows:

$$\text{armour and first wall} = fC_{FW}P_{fusion}M_{FW} \qquad 48$$

$$\text{blanket} = fC_{blanket}P_{fusion}(1 - e^{-aM_{blanket}}) \qquad 49$$

$$\text{shield} = fC_{shield}P_{fusion}M_{shield}e^{-dL_{blanket}}e^{-fL_{shield}} \qquad 50$$

$$\text{coils} = fC_{coils}P_{fusion}M_{coils}e^{-bL_{blanket}}e^{-c(L_{shield}+L_{vv})} \qquad 51$$

where $P_{fusion}$ is the fusion power, $M_{FW}$ is the mass of the first wall and its armour, $M_{blanket}$ is the mass of the blanket, $M_{coils}$ is the mass of the TF and PF coils (heating in the CS is negligible), $L_{blanket}$, $L_{shield}$ and $L_{vv}$ are the integrated line densities of armour + first wall + blanket, the shield, and the vacuum vessel respectively. Although the neutronics calculations give absolute results, we have introduced a factor *f* to renormalize the power deposition making the assumption that 100% of the neutrons are absorbed.

## 7. Heat extraction from first wall and electricity generation

The heat from the primary coolant is transferred to a secondary coolant (working fluid), which is used to generate electricity in a power conversion cycle. The theoretical efficiency of conversion will be determined by the mean temperature at which heat is added to the working fluid by the primary coolant, and the temperature at which heat is rejected to the environment. However, the primary coolant temperature must be sufficiently low for the materials in the reactor structure to be below their practical limits. It is further necessary to minimise the power demand for pumping the primary and secondary coolants. Two models are available that combine the thermohydraulics of the blanket and first wall with the secondary cycle: a simple model and a detailed model, described below. PROCESS assumes that for a pulsed reactor the energy conversion system can be rapidly switched between operation and standby modes, without losing any additional thermal or electrical energy. The dwell time between pulses is input by the user.

### 7.1. Simple model

In this model, the power required to pump the primary coolant through each of the first wall, breeder zone, divertor, and shield, is calculated as a user input fraction of the thermal power deposited in the coolant from the reactor. The pumping power is deposited in the primary coolant. The default values for this fraction are 0.0005 for a water coolant, and 0.085 for a helium coolant, based on very preliminary calculations made for DEMO in EUROfusion. The separation of power into first wall, breeder zone, divertor, and shield allows a different coolant to be assumed for each system if required.

The gross electric power is derived using a thermal efficiency based on the user's choice of blanket. The options are water-cooled lithium-lead (WCLL), helium-cooled lithium-lead (HCLL), or helium-cooled pebble-bed (HCPB). The resulting thermal efficiencies used are taken from studies that modelled Rankine cycles for the different options of a helium-cooled primary circuit with a top temperature of 500ºC (6), and a water-cooled primary circuit with a top temperature of 320ºC (7). (For historical reasons in both cases the divertor was cooled by water with a top temperature of 150ºC in the helium-cooled reactor, and 250ºC for the water-cooled reactor.) Hence, no variation of efficiency with primary coolant temperature is possible using the simplified model; indeed, no temperatures are even considered in the model. The defined thermal efficiencies for the given blanket choices are shown in Table 3. For a helium-cooled reactor a penalty is applied as the coolant in the divertor has to operate at much lower temperature than the blanket, which may be the case because of the greater heat flux that has to be removed. Efficiencies differ depending on whether the heat from the divertor is utilised to preheat the secondary coolant, or is discarded as waste heat. Note that these thermal efficiencies are for the cycle only, describing the conversion of primary heat to gross electric power. The overall plant net power, accounting for recirculating power (including, for instance, the primary coolant pumping power demands), and hence the plant net efficiency, will be lower.

Table 3. Simple energy conversion model: secondary cycle thermal efficiency, defined as gross electric power divided by thermal power deposited in the secondary coolant. *f* = fraction of heat to the divertor

| Primary Coolant | Water | | Helium | |
|---|---|---|---|---|
| Divertor heat used | Yes | No | Yes | No |
| Efficiency | 31% | n.a. | $0.411 - 0.339\,f$ | 41.1% |

## 7.2. Detailed model: first wall

The detailed model for heat extraction and power conversion calculates the maximum temperature of the first wall for given coolant inlet and outlet temperatures and channel dimensions. An iteration loop is utilised to decrease the thickness of the first wall, up to practical limits, if the temperature is found to be above the material limits. From this, the efficiency of the secondary cycle can be found from the coolant outlet temperature. The divertor heat is included in the primary cycle.

The outlet temperature of the primary coolant is (a) for water, 20 K below the boiling point; (b) for helium, a user input.

To calculate the maximum temperature of the first wall, the method of LeClaire is used (8), (9). In this approach, the first wall is assumed to consist of a set of parallel pipes, such that the surface is not flat, but a repeating semi-circular pattern. This geometry is unlikely to be used in practice, but allows a convenient analytical approach. The calculation of the peak temperature assumes that this occurs at the point closest to the plasma on each pipe and where the coolant temperature is at its maximum (which occurs at the top of the front face of the module). The temperature is calculated analytically by considering the conduction of heat through the pipe structure to the coolant, where heat is deposited both volumetrically from the incident neutron power, and upon the surface from radiative power. The heat transfer coefficient is calculated using the Sieder-Tate correlation. If the first wall temperature is found to be above specified limits, the thickness of the first wall is reduced, until it becomes too low to withstand the pressure of the coolant. (Note that if both criteria are satisfied using the initial inputs, the code does not vary the thickness of the first wall.) If the temperature and thickness requirements cannot both be satisfied the code returns an error. The peak hoop stress is given by Lamé's solution for a thick-walled cylinder:

$$\sigma_p = \frac{P(b_{fw}^2 + a_{fw}^2)}{(b_{fw}^2 - a_{fw}^2)} \qquad 52$$

where $b_{fw}$ and $a_{fw}$ are the outer and inner radii of the first wall pipes, and P is the maximum coolant pressure. This must be less than the permissible stress. The plasma-facing side of the pipe may be eroded by sputtering, so an *ad hoc* adjustment is made,

$$\sigma_p = \frac{P((b_{fw} - w_{erosion})^2 + a_{fw}^2)}{((b_{fw} - w_{erosion})^2 - a_{fw}^2)} \qquad 53$$

where $w_{erosion}$ is the specified erosion thinning over the lifetime of the first wall. There is also a neutron fluence limit which determines the lifetime of the first wall and blanket – see section 8.1 below.

The power required to pump the coolant through the first wall and breeder zone is not trivial to calculate, as it depends sensitively on the diameter of the channels, and on the design of the feeder pipes and manifolds. The relevant algorithms are still under development.

## 7.3. Detailed model: energy conversion

From the coolant outlet temperature, the thermal efficiency of the power conversion cycle is determined. The user can choose between a steam Rankine cycle and a supercritical carbon dioxide Brayton cycle.

If the Rankine cycle is chosen and the primary coolant is water, it is assumed that the cycle is similar to that of pressurised water reactors currently in operation. This cycle was modelled for a range of different top temperatures in order to find a correlation of cycle efficiency with temperature. The modelling method is described in (7). A penalty of 0.042 was subtracted from the efficiency to account for pressure losses in the cycle using a more detailed model as a benchmark (7). If the Rankine secondary is chosen but the primary coolant is helium, it is assumed that the cycle is a superheated-steam Rankine cycle. The results of modelling by Dostal (10) were used, but with an efficiency penalty of 0.0179 to give agreement at one point with a benchmark (6). For the supercritical $CO_2$ cycle, the correlation of efficiency with temperature is derived from results of cycle modelling carried out by CCFE in collaboration with industry. The derived fits are in Table 4. For both Rankine cycles the divertor heat is used in a separate heat exchanger to preheat the feedwater, while for $CO_2$ cycle it is used in the main heat exchanger. In both cases the divertor heat is counted as primary heat, and is included in the calculation of the efficiency.

Table 4. Fitting functions for secondary cycle efficiency. $T_1$ and $T_3$ are the maximum temperatures of the primary blanket and divertor coolants; $T_2$ is the temperature of the secondary fluid at the inlet to the turbine. $\Delta\eta$ is a correction to represent the loss in cycle efficiency due to the lower temperature of the divertor coolant.

| Secondary cycle | Primary coolant | Efficiency ($T_2$ in °C) | $T_2$ range (°C) | $T_1$ - $T_2$ (°C) | $T_3$ (°C) |
|---|---|---|---|---|---|
| Steam Rankine | Water | $0.3720 \ln(T_2 + 273) - 2.0219$ | 275-310 | 40 | 250 |
| | Helium | $0.1802 \ln(T_2 + 273) - 0.7823 - \Delta\eta$ | 384-642 | 20 | 150 |
| Supercritical $CO_2$ Brayton | Water or Helium | $0.4347 \ln(T_2 + 273) - 2.5043$ | 135-750 | 20 | $T_3 = T_2$ |

## 8. Availability

The availability of a power plant is crucial for generating electricity economically. A new availability module has recently been added. Our definitions are as follows. *Availability* is the fraction of the time in which the plant is operating normally. For a pulsed reactor the dwell time in between pulses is considered to be normal operation. *Capacity factor* is the electrical energy delivered to the grid over the lifetime of the plant, divided by the maximum rate at which electrical power can be delivered. These are the same for a steady-state, fixed power reactor, but for a pulsed reactor capacity factor will be less than availability. (No allowance has been made for load following – reduction of output at times of low demand.)

The total availability is derived from the addition of the planned and unplanned availabilities, $U_{planned}$ and $U_{unplanned}$, and a term to take account of the overlap:

$$Availability = A_{tot} = 1 - (U_{planned} + U_{unplanned} - U_{planned}U_{unplanned}) \qquad 54$$

The cost of electricity (COE) is the mechanism by which the availability is fed back into the PROCESS optimiser. For a pulsed reactor the duty cycle $F_{dc}$ is calculated using the pulse length and the time between pulses (dwell time). The capacity factor is $A_{tot}F_{dc}$. The code uses this value and cost data (including the capital cost) to estimate the cost of electricity for the plant.

The lifetime of the components of a pulsed reactor may be substantially reduced because of fatigue, but this is not currently taken into account.

### 8.1. Planned Unavailability

The planned unavailability in PROCESS is linked to the lifetimes of the blanket and divertor and the time taken to replace them. The lifetime for the blanket is based only on the neutron flux, using the following very loose scaling:

$$t_{life,blanket} = \frac{\phi_{blanket}}{q_{blanket}} \qquad 55$$

where $\phi_{blanket}$ is the allowable fast neutron fluence and $q_{blanket}$ is the nominal neutron wall load, defined as the fusion power divided by the first wall area. In contrast, the divertor lifetime is estimated using the particle and photon heat load:

$$t_{life,div} = \frac{\phi_{div}}{q_{div}} \qquad 56$$

where $\phi_{div}$ is the allowable cumulative heat load and $q_{div}$ is the peak heat load on the divertor. This formula will usually ensure that the divertor lifetime reduces as the power into the divertor increases, but the absolute values derived should not be taken seriously.

PROCESS calculates which of the two components has the shorter lifetime. For example, if the divertor has a shorter life, one or more outages, $n_{outages}$, may be required for divertor replacement within the lifetime of a single blanket. The unavailability is given by:

$$U = \frac{t_{main}}{t_{op} + t_{main}}, \qquad 57$$

where $t_{op}$ is the total operational time, and $t_{main}$ is the total time required for maintenance, both over the entire life of the machine. The operational time $t_{op}$ is given by $t_{life,blanket}$ (or $t_{life,div}$, if lower). The total planned unavailability is then:

$$U_{planned} = \frac{n_{outages} t_{div,repair} + t_{complete,repair}}{t_{life,blanket} + n_{outages} t_{div,repair} + t_{complete,repair}}, \qquad 58$$

where $t_{div, repair}$ is the time to replace the divertor and $t_{complete, repair}$ is the time taken to replace both the blanket and the divertor.

The time to replace the blanket and divertor are estimated by Crofts et al (11), who studied the influence of the number of remote handling systems on the length of scheduled maintenance. A fit to their results gives the time to repair both the blanket and divertor as:

$$t_{repair}(months) = \frac{21}{N^{0.9}} + 2 \qquad 59$$

where $N$ is the number of remote handling systems working in parallel. The extra two months are to allow the dose-rate to reduce to an acceptable level before remote handling operations start, and to allow pump-down and preparation for operation at the end of the shutdown. Crofts et al comment that to replace the blanket one must remove the divertor also. On the other hand it is possible to replace the divertor alone and this is estimated to take 70% of the time taken to repair the blanket.

## 8.2. Unplanned downtime

Each subsystem is represented by a simple model that tries to capture the degradation of reliability when approaching operational and technological limits. This increases the risk of unplanned downtime as the design margins are reduced. This approach is well suited to a systems code, as it ensures that the optimiser will choose a design point that has adequate design margins whenever possible. The total unplanned unavailability is the sum of the unplanned unavailabilities for each system.

For the heating and current drive systems the unplanned unavailability is taken as 2%. This is far less than currently operating systems - large improvements will need to be made in order to meet the requirements of a power plant. The failure rate for the steam turbine system can be estimated from experience (12) as $9.39 \times 10^{-5}$ failures per hour, with an associated average repair time of 96 hours.

### 8.2.1. Magnets

It is likely that the chance of a quench in a magnet is the largest driver of the risk of unplanned unavailability, and this may depend on the temperature margin in the TF coils – the difference between the actual temperature and the critical temperature of the superconductor (section 2.2). The unplanned unavailability of the magnet system is given by:

$$U = \frac{\tau_{main}}{\tau_{main} + \tau_{bquench}}, \qquad 60$$

where $\tau_{main}$ is the maintenance time for the magnet system and $\tau_{bquench}$ is the estimated time between quenches of the magnet system. This is calculated as follows.

$$\tau_{bquench} = \begin{cases} L_{Mag} & T_{marg} > \dfrac{T_{marg,min}}{c} \\ L_{Mag} \dfrac{T_{marg} - T_{marg,min}}{T_{marg,min}} \dfrac{c}{1-c} & T_{marg,min} < T_{marg} < \dfrac{T_{marg,min}}{c} \\ 0 & T_{marg} < T_{marg.min} \end{cases} \qquad 61$$

where
$L_{Mag}$ = lifetime when design margin is large,
$T_{marg}$ = superconductor temperature margin,
$T_{marg,lim}$ = temperature margin lower limit,
$c$ = determines the temperature margin at which lifetime starts to decline.

The suggested range for $c$ is 0.9-0.99. Figure 11 shows an example for a minimum temperature margin of 1.5 K. The magnet never quenches if the temperature margin is above a critical value, and it will not operate at all below the temperature margin lower limit. In between, there is a finite risk of quench.

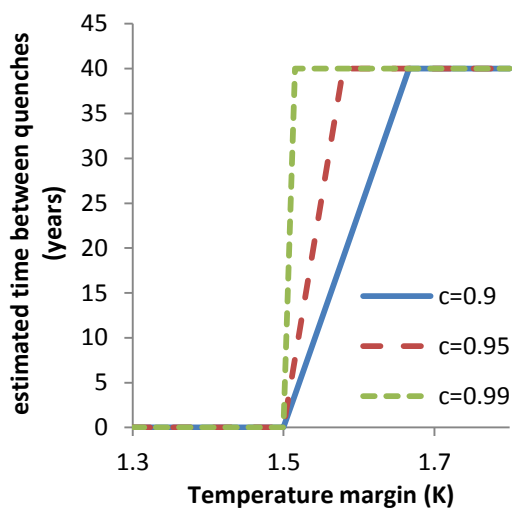

Figure 11. Illustration of the approach used to estimate unplanned failures: estimated time between quenches in the TF coils

8.2.2. First wall, blanket and divertor

The first wall and blanket will be subject to neutron irradiation during operations. In addition, the first wall will be exposed to photon radiation, and particle bombardment by ions and neutral atoms. The unplanned downtime is based on the number of cycles a blanket set experiences before replacement. (This model is restricted to pulsed reactors). The number of cycles between planned blanket replacements, $N$, is determined by the blanket lifetime which is based on the neutron flux (section 8.1 above):

$$N = \frac{t_{life,blanket}}{t_{cycle}} \qquad 62$$

This approach allows PROCESS to improve the availability of the blanket by *increasing* the neutron flux, until the planned downtime starts to dominate. This is counter-intuitive, but is correct if one assumes that the neutron flux affects the total life of the blanket but not the cycle life.

The life of the blanket is expressed in terms of a reference number of cycles $N_{ref}$. The probability of failure in one pulse cycle before the reference cycle life is a constant, $p_f$. During the reference lifetime the instantaneous availability after $n$ cycles since the blanket was last repaired or replaced is:

$$a(n) = a_0 = \left(1 - \frac{p_f t_{main}}{t_{cycle}}\right) \qquad n \leq N_{ref} \qquad 63$$

where $t_{main}$ is the time required to repair the blanket, $t_{cycle}$ is the length of one pulse cycle. After the reference lifetime we assume that the reliability of the blanket starts to decline, so the instantaneous availability is given by:

$$a(n) = a_0 \left(\frac{N_U - n}{N_U - N_{ref}}\right) \qquad n > N_{ref} \tag{64}$$

where $N_U$ is the cycle when the blanket fails with 100% probability. The availability decreases linearly beyond the reference lifetime. Integrating the instantaneous availability gives the mean availability over the planned cycle life $N$:

$$\begin{aligned} N \leq N_{ref} \quad & A(N) = a_0 \\ N > N_{ref} \quad & A(N) = \frac{a_0}{N_U - N_{ref}}\left(N_U - \frac{N_{ref}^2}{2N} - \frac{N}{2}\right) \end{aligned} \tag{65}$$

The availability of the divertor is estimated in a similar way.

### 8.2.3. Vacuum System

The vacuum system will be extensive and complex, as it must capture unburnt fuel, helium ash and impurities, as well as evacuating the reactor before operation. PROCESS assumes that there is a pumping duct between every pair of adjacent TF coils, and that cryopumps are used, so that for each duct there are two pumps that can be regenerated alternately. During planned maintenance broken pumps can be replaced, so the calculation of unplanned downtime is done for an operational period between planned shutdowns and then multiplied by the number of operational periods in the machine lifetime. The total operational time between shutdowns is

$$t_{op} = \frac{t_{op,total}}{N_{shutdowns} + 1} \tag{66}$$

where $t_{op,total}$ is the total operational time in the life of the machine. The failure rate for a cryopump is taken from (13):

$$p_f = 2 \times 10^{-6} \text{ /hour} \tag{67}$$

If the total number of pumps is $N_{tot}$, then the probability of $n$ failures in the operational period $t_{op}$ is:

$$P(n) = \binom{N_{tot}}{n} (t_{op}p_f)^n \left(1 - t_{op}p_f\right)^{N_{tot}-n} \tag{68}$$

where $\binom{N_{tot}}{n}$ are the binomial coefficients. If the number of failures exceeds the number of redundant pumps then it will cause additional unplanned downtime. The total downtime over the entire operational period is then

$$t_{down} = (N_{shutdowns} + 1)t_{main} \sum_{n=N_r+1}^{N_{tot}} P(n)(n - N_r) \tag{69}$$

where $t_{main}$ is the unscheduled maintenance time for a vacuum pump, and $N_r$ is the number of redundant pumps. If there are several redundant pumps then this unscheduled downtime can be reduced to a negligible level. Then the unplanned unavailability is

$$U_{vac} = \max(U_{min}, \frac{t_{down}}{t_{op,total} + t_{down}}) \tag{70}$$

The lower limit $U_{min}$ allows for common mode failures that affect several pumps.

## 9. Application to DEMO

A model for a pulsed reactor generating 500 MW net electricity has been obtained using PROCESS, referred to as DEMO A (1). Some of the engineering aspects are discussed here, and illustrated in Figure 1. The engineering constraints selected are listed in Table 5.

Table 5. Some of the engineering constraints applied in the DEMO A model.

| Limiting constraints | Limit applied |
|---|---|
| Current density in central solenoid at end of flat-top = 0.25×critical current density | < 1.36×10$^7$ A/m$^2$ |
| Current density in winding pack of TF coils = 0.5×critical current density | < 3.67×10$^7$ A/m$^2$ |
| Stress in the case of the TF coil (von Mises stress) | 660 MPa |
| Conductor temperature in quench of TF coil | < 150 K |
| Thickness of conduit of TF coil conductor | > 4 mm |
| Ratio of power crossing the separatrix to plasma major radius (Psep/R) (MW/m) | 17 MW/m |
| Net electric power output | > 500 MWe |
| **Constraints applied but found not to be limiting** | |
| Voltage generated in quench of TF coil | < 20 kV |
| Minimum availability value | >75% |
| Nominal neutron wall load | < 8 MW/m$^2$ |
| **Constraints described in this paper but not applied** | |
| Central solenoid temperature margin lower limit | |

The machine is large - the outside dimensions of the cryostat are 36 m (diameter) x 29 m (height). The TF coils allow the neutral beam to be tangent to the plasma axis or even further out, allowing optimum current drive to be achieved if required. The power flows are summarised in Table 6, and some thermodynamic parameters are in Table 7. The 40% wall plug efficiency assumed for the heating system contrasts with achieved values of about 25% (14).

Table 6. Power flows.

| Power Balance for Reactor | MW |
|---|---|
| Fusion power | 1686 |
| Power from energy multiplication in blanket and shield (MW) | 321 |
| Injected power (MW) | 50 |
| Ohmic power (inductive power transfer to plasma) (MW) | 1 |
| Power deposited in primary coolant by pump | 14 |
| Total | **2072** |
| Heat extracted from armour and first wall | 435 |
| Heat extracted from blanket | 1297 |
| Heat extracted from shield | 2 |
| Heat extracted from divertor | 338 |
| Total | **2073** |

Table 7. Thermodynamic and energy parameters.

| Wall plug efficiency of neutral beam injection system | 40% |
|---|---|
| Primary coolant | Helium |
| Primary coolant inlet/outlet temperatures (blanket) | 300 ºC / 550ºC |
| Primary coolant inlet/outlet temperatures (divertor) | 100/150°C |
| Secondary coolant | Water/steam |
| Gross electric power* / high grade heat (*Power for pumps in secondary circuit already subtracted) | 36.8 % |
| Net electric power / total nuclear power | 24.9 % |

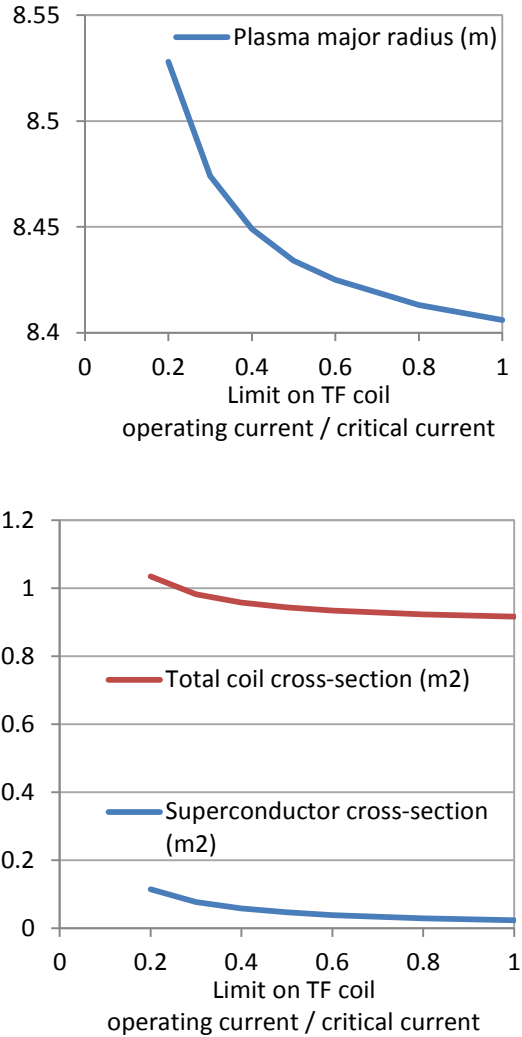

Figure 12. Dependence of plasma major radius, total TF coil cross-section and superconductor cross-section on normalised maximum permissible current density in TF coil. Figure of Merit is major radius.

Figure 12 illustrates the way in which multiple constraints can conspire to limit the space for variability of the optimised model. Even when the ratio of operating current to critical current in the TF coils is increased by a factor of five, the total coil cross-section decreases only a little, because of the large non-superconducting area needed for copper stabiliser, insulation, and, especially, structural support. The result is that the plasma major radius reduces by just 12 cm, or 1.4%. This shows that to make effective use of improved superconductors, stronger structural materials would also be required.

Another illustration of this effect of multiple constraints is shown in Figure 13. As the output power requirement is increased, the code increases the seeded impurity fraction, in order to increase the radiation fraction, thus maintaining the power per unit length into the divertor zone at a constant value. The injected power was fixed in this run, so the electric power required for current drive drops as a fraction of gross electric power. As the machine gets bigger, the increased space allows the toroidal field to increase slightly. Because of these factors and others, the major radius grows by only 9%, even though the power output has increased by a factor of 2.2.

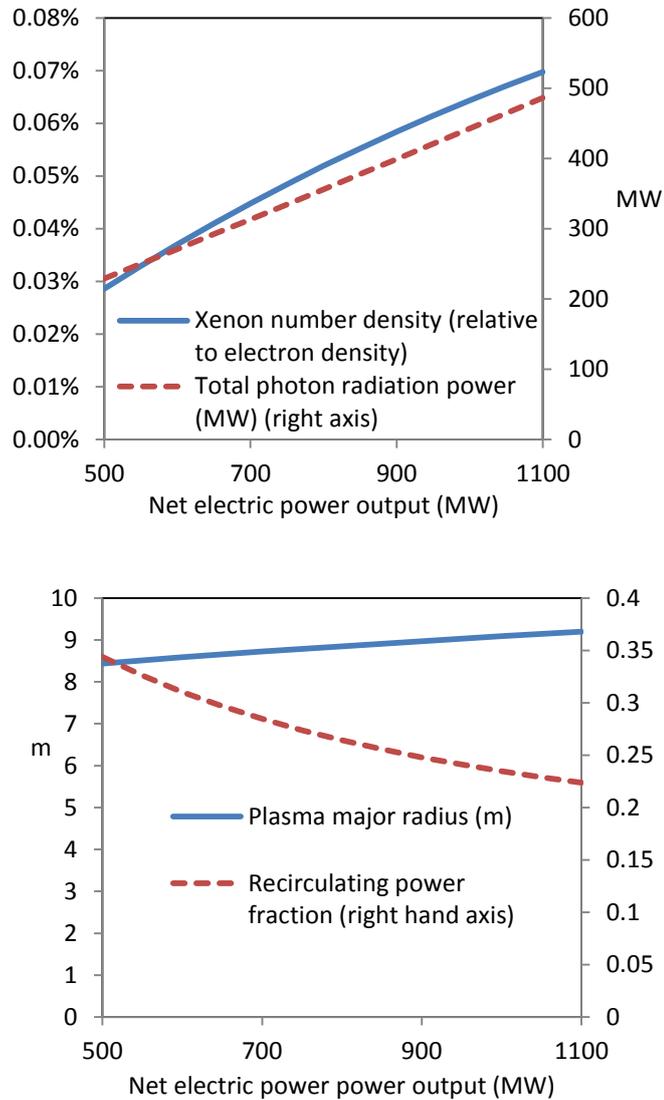

Figure 13. Dependence of the density of the seeded impurity (xenon), radiation power, major radius and recirculating electric power on net electric output required. Figure of Merit is major radius.

Fully annotated input and output files for these models are available in the Supplementary Data accompanying this paper.

## 10.     Discussion and Future work

A new neutronics module is being developed (15). The time-dependent code FATI was used to simulate the depletion of breeding and neutron multiplying isotopes in an HCPB blanket. 594 models were run, covering the parameter space of lithium fraction, $^6$Li enrichment, and blanket thickness. The results were fitted with analytical expressions, which will be incorporated into PROCESS. In the future it will be important to obtain improved parametric expressions for energy multiplication, and for nuclear energy deposition in the coils, with and without neutral beam ducts. New estimates of the approximate likely cost of building a tokamak fusion power station, on the assumption that the outstanding issues can be resolved, are being developed for PROCESS.

PROCESS allows the user to choose which constraints to impose and which to ignore, so when evaluating the results it is vital to study the list of constraints used. Work is underway on the sensitivity of PROCESS results to variations in input parameters, and on the robustness of the optimiser in finding global solutions. New algorithms submitted by collaborators can be incorporated – for example safety, first wall erosion, and fatigue life will be crucial and are not yet taken into account. The PROCESS homepage is [www.ccfe.ac.uk/powerplants.aspx](www.ccfe.ac.uk/powerplants.aspx).

## 11. Acknowledgments

We would like to thank H. Latham for the Thermoflow modelling, J. Shimwell and P. Pereslavtsev for the neutronics model, and C. Bachmann, F. Maviglia, and E. Surrey for helpful comments. This work has been carried out within the framework of the EUROfusion Consortium and has received funding from the Euratom research and training programme 2014-2018 under grant agreement No 633053 and from the RCUK Energy Programme [grant number EP/I501045]. To obtain further information on the models underlying this paper please contact the authors, or PublicationsManager@ccfe.ac.uk. The views and opinions expressed herein do not necessarily reflect those of the European Commission.